\documentstyle[12pt]{article}
\begin{document}
\def\hb{\hfill\break}
\def\sgn{\hbox{\rm sgn}}


\title{ The EPR correlations and the chameleon effect }
\author{Luigi Accardi,  Massimo Regoli }
\maketitle

\centerline{\it Centro Vito Volterra, Universit\`a degli Studi di Roma
``Tor Vergata''}

\centerline{\it email: accardi@volterra.mat.uniroma2.it,
WEB page: http://volterra.mat.uniroma2.it}

\begin{abstract}
We describe an experiment in which two non communicating computers, starting
from a common input in the form of sequences of pseudo--random numbers in the
interval $[0,2\pi]$, and computing deterministic $\{\pm 1\}$--valued
functions, chosen at
random and independently, produce sequences of numbers whose correlations
coincide with the EPR correlations and therefore violate Bell's inequality.

The experiment is the practical implementation of a mathematical model
of a classical, deterministic system whose initial state is chosen at
random from its state space, with a known initial probability
distribution, and whose dynamics exhibits the chameleon effect described
below. Such a system satisfies the constraints of pre--determination,
locality, causality, local independent choices, singlet law and
reproduces the EPR correlations.
\end{abstract}

\vfill\eject

\section{ Introduction }

Since the appearence of Bell's paper \cite{[Be64]} in 1964 the following
problem has played a central role in the debate on the foundations of
quantum mechanics:

Do there exist classical systems which reproduce the EPR (or singlet)
correlations under the constraints of pre-determination,
locality principle, singlet condition and causality principle? (cf.
Section (2) below for comments on the meaning of these terms).

The conceptual implications of this, apparently technical, question
reside in the fact that the possibility (also called a hidden variable
theory) of a classical interpretation of natural phenomena (e.g. the EPR
type experiments), in which the properties of the physical objects are
pre-determined, and not created at random by the act of measurement,
is often identified to the possibility of having a realistic
point of view on the natural phenomena. For this reason the above question
is usually considered equivalent to the following:

Is it possible to simultaneously maintain: (i) a realistic point of view
on the natural phenomena; (ii) the predictions of quantum mechanics (EPR
correlations); (ii) the locality (or causality) principle?

It is a fact that the (practically) unanimous answer to this question, in
the past 37 years, has been a flat: No!

The argument, supporting this unanimous answer, is that:

(i)  the EPR correlations violate Bell's inequality

(ii) Bell's analysis proves that any system, satisfying pre-determination,
locality and causality cannot violate Bell's inequality.

This argument is often condensed in the statement that there cannot
exist local realistic hidden variable theories.

The quantum probabilistic point of view \cite{[Ac81a]}, challanged
Bell's analysis on two different fronts namely:

1) by proving that it is theoretically inconclusive, i.e. that statement
(ii) above is wrong (cf. \cite{[AcRe00b]} for a survey of these arguments)

2) by constructing a general mechanism (called ``chameleon effect'')
\cite{[Ac93]} which shows how to construct classical
systems violating the hidden mathematical assumption, implicit in
the proof of Bell's inequality.

Yet, until recently, no attempt had been made to solve this controversy
by means of an experiment. In other words, until recently, the general
mechanism mentioned in item (2) above had not been substantiated into an
example of a concrete classical physical system which satisfies
pre-determination, locality and causality and yet violates Bell's inequality.

It is true that the list of experiments, some of which truly spectacular
\cite{[GrHoZe93]},
confirming the emergence of the EPR correlations in several quantum
mechanical phenomena is nowadays very long and continuously increasing.

However, being performed on systems with a strong quantum behavior,
these experiments, while confirming the important statement that singlet
correlations effectively appear in nature, cannot say anything on the
main thesis of Bell, namely that no classical system can reproduce these
correlations under the above constraints.

The first experiment in this direction was performed in april 1999
and reported in \cite{[AcRe99a]}, \cite{[AcRe99b]}.
Its results supported the point of view of quantum probability, in the
sense that a local realistic violation  of Bell's inequality was detected,
however without reprodu\-ction of the EPR correlations.

The second experiment in this direction was performed in june 2001 and
reported in \cite{[Ac01a]}, \cite{[Ac01b]}. In it the EPR correlations
were faithfully reproduced by local, independent and even deterministic,
macroscopic, classical systems. As explained above, violation of Bell's
inequality is a consequence of these correlations.

The third experiment will be performed publicly as a satellite
activity of the ``Japan-Italy
Joint workshop on Quantum open systems and quantum measurement'',
(Waseda University, 27--29 September 2001) and consists of a
sophisticated elaboration of the second one, based on three separated
and non communicating computers. Thus, in this case, the classical
macroscopic systems, violating by local independent choices the Bell's
inequality, will be the computer themselves and the  persons operating
them. In this sense, although involving computers, the one described
below is not a simulation of an experiment, but a real one.

In the following we describe the simple idea on which this experiment is
based. To this goal, let us first of all state the problem more precisely.\bigskip

{\bf Acknowledgements}\medskip

The authors are grateful to Kentaro Imafuku for several interesting
discussions which led to an improvement of the presentation of the
material discussed below.

\section{  Predetermination, causality, locality and the chameleon effect}

Consider a classical dynamical system composed of two particles $(1,2)$.
The term classical here means that

-- each of the particles has a state space $S$

-- observables are functions on $S$:
$$
\tilde S^{(1)}_a, \tilde S^{(2)}_a : S \to \{ \pm 1 \}
$$

-- the measured values of the observables depend only on the initial state
and on the dynamics (of each individual system).

The third requirement is the pre--determination requirement on which both
EPR and Bell strongly insist as a crucial requirement of a realistic point
of view. A crucial remark here is that the term ``pre--determination'' can
be interpreted in the usual way of classical statistical mechanics: the
value of the observable (e.g. the color of a ball) ``is there'' and we
passively register it; or in an active way, like the color of a
chameleon, which is ``pre--determined'' to become green on a leaf and
brown on a piece of wood. This is why one should distinguish between
``Einstein (or ballot--box) realism'' and ``chameleon realism''.

The principle of causality asserts that:

-- the state of any system at time $0$ (present) is independent of the state
of any other system at time $T>0$ (future)

This means that:

-- at time $t=0$ the particles do not know which measurement will be
done on them

and, mathematically it is translated in the fact that, at time $0$, the global
( i.e. system--apparatus) state $\psi$ should be factorized as follows:
\begin{equation}\label{causl}
 \psi=\psi_{system}\otimes \psi_{apparatus}
\end{equation}

The locality principle, both in the EPR's and in Bell's words, means that:

``... the result $B$ for particle $2$ does not depend on the setting $a$,
of the magnet for particle $1$, nor $A$ on $b$....'' \cite{[Be64]}.

This means first of all that the two apparata make independent choices,
i.e. that their preparations at time $0$ are independent:
\begin{equation}\label{locapp}
\psi_{apparatus}=\psi_{apparatus \ 1}\otimes \psi_{apparatus \ 2}
\end{equation}
and, moreover, that two far away particles:

-- don't feel their mutual interaction

-- don't feel any interaction with a far away measurement apparatus

Obviously this does not exclude that:

``... the result $B$ {\bf for particle $2$} depends on the setting $b$,
of the magnet {\bf for particle $2$}, and $A$ on $a$....''

This statement expresses the essence of the {\bf chameleon effect}.
In fact, if we consider the two ``chameleon observables'': ``color on the
leaf'' and ``color on the wood'', in analogy to ``spin in direction
$a$'' and
``spin in direction $b$'', we see that the dynamical evolution of the
chameleon, who becomes green when approaching the leaf and brown when
approaching the wood, depends on the observable we measure.

More precisely (notice that the definition below is valid both for
classical and quantum systems):\bigskip

\noindent{\bf Definition}. Let $I$ be a set and, for each $x\in I$, let
be given an observable $\tilde S_x$ of a system $\sigma$ with state space $S$.
The system $\sigma$ is said to realize the chameleon effect with respect to
the observables $\tilde S_x$ if: whenever the observable $\tilde S_x$ is
measured, the dynamical evolution (discrete time) of the system
$$
T_x:S\to S
$$
depends on the measured observable $\tilde S_x$.\bigskip

\noindent{\bf Remark}.
In our experiments the indices $x$, $y, \ldots$ are vectors in the unit
circle in ${\bf R}^2$ (or angles in $[0,2\pi ]$).

The dynamics of the full system:

(particle $1$, particle $2$, apparatus $1$, apparatus $2$),

is unitary (an automorphism, at Heisenberg level). We will consider the
reduced dynamics (cf. the Appendix), in Heisenberg representation, of the
subsystem:

$(1,2) =$ (particle $1$, particle $2$)

which in general is not an automorphism, but a completely positive
map (for classical systems this is equivalent to positivity).

The observables of the system $(1,2)$ are functions on $S\times S$ and
we denote $f\otimes g$ those special observables of the form
$$f\otimes g (u,v) =f (u) g(v)\qquad ;\qquad u,v\in S$$
where $f, g$ are functions on $S$. A local (reduced) dynamics for the
system $(1,2)$ is a positive map $P$, of the space of functions
on $S\times S$ into itself, with the property that, for any two
functions $f, g$ on $S$,
\begin{equation}\label{locdyn}
 P(f\otimes g)=P_1(f)\otimes P_2(g)
\end{equation}
where $P_1,P_2$ are positive maps of the space of functions on $S$
into itself. This means that particle $1$ and particle $2$ evolve
with their own dynamics independently of each other. If
condition (\ref{locdyn}) is satisfied, we will write
\begin{equation}\label{locdyn2}
 P=P_1\otimes P_2
\end{equation}

For example, let $T_1,T_2 : S \to S$ be two reversible dynamics
(invertible maps)
and let $\tau_1,\tau_2 : S \to {\bf R}$ be two positive maps. Then the map
\begin{equation}\label{locdynex}
PF(u,v) := \tau_1(u) F(T_1u,T_2v) \tau_2(v) \quad ; \ u,v\in S
\end{equation}
satisfies the locality condition (\ref{locdyn}) with
\begin{equation}\label{p1p2}
P_1(f)(u):= \tau_1(u) f(T_1u) \quad ,\quad P_2(g)(v):= g(T_2v) \tau_2(v)
\end{equation}
here $F$ is a function on $S\times S$ and $f, g$ functions on $S$.
The reduced dynamics in our experiment will be of this form.

Given a local dynamics $P$ and a state $\psi$, the Schr\"odinger
evolution is defined by
\begin{equation}\label{schrd}
\psi \mapsto \psi \circ P
\end{equation}
and, since for any observable $F$ one has
\begin{equation}\label{dual}
 \psi \circ P (F) = \psi (P(F) )
\end{equation}
it defines the Heisenberg evolution by
\begin{equation}\label{heis}
 F \mapsto P(F)
\end{equation}
The two representations are equivalent.

\section{  Physical idea of the experiment: general structure}

We will choose the state space $S$ (hidden variables) to be the
interval $[0,2\pi ]$ (identified to the unit circle in the plane) and
we will construct:

-- for any pair of angles $a,b$, in $[0,2\pi ]$, two functions (observables)
\begin{equation}\label{pm1val}
\tilde S^{(1)}_a, \tilde S^{(2)}_b : S \to \{ \pm 1 \}
\end{equation}

-- for any pair of angles $a,b$, in $[0,2\pi ]$, two local (single particle)
dynamics (local determinism) $P_{1,a}, P_{1,b}$

-- a classical probability distribution $\psi_o$ on the state space $S\times S$
(initial distribution of the hidden variables)

\noindent with the following properties: for any pair of angles $a,b$, in $[0,2\pi ]$,
the correlations of the two random variables
$ \tilde S^{(1)}_a,\tilde S^{(2)}_b$,
with respect to the measure $\psi_o$, evolved with the reduced dynamics
$P:=P_{1,a} \otimes P_{1,b}$ (cf. (\ref{locdyn})), are exactly the
singlet (EPR) correlations, i.e.
\begin{equation}\label{eprcorsh}
\int\int
\tilde S^{(1)}_a (\mu_1) \tilde S^{(2)}_b(\mu_2)
(\psi_o \circ P)(d\mu_1,d\mu_2)
=:\langle S^{(1)}_aS^{(2)}_b\rangle = - cos (b-a)
\end{equation}
Using (\ref{dual}) and (\ref{locdyn}) we can write the correlations
(\ref{eprcorsh}) in Heisenberg representation:
\begin{equation}\label{eprcorhe}
\int\int
P_{1,a}( \tilde S^{(1)}_a)(\mu_1)
P_{1,b}(\tilde S^{(2)}_b)(\mu_2)\psi_o(d\mu_1,d\mu_2)
\end{equation}

In order to complete our construction we want to introduce in our
classical dynamical system the analogue of the singlet condition.
In the standard (quantum) situation the singlet condition expresses the
law of conservation of spin: the property of having spin zero is a
constant of motion of the system.

Our classical analogue of the singlet condition will be the constraint
that at time $t=0$ (initial time) the state $\mu^0_1$, of particle
$1$, and the state $\mu^0_2$, of particle $2$, coincide:
$$
\mu^0=\mu^0_1=\mu^0_2
$$
This means that the initial state space is not the whole phase space
$S\times S$, but the surface (in $S\times S$) defined by the equation:
\begin{equation}\label{singlsurf}
\mu_1 -  \mu_2 = 0
\end{equation}
If we identify the unit circle in the plane to the interval $[0,2\pi ]$, this
surface is simply the diagonal of the square $[0,2\pi ]\times [0,2\pi ]$.

We do not impose other constraints and we suppose that, on the initial
surface (\ref{singlsurf}), all the cells of the phase space with equal
Lebesgue measure are equiprobable. Therefore our initial probability measure
$\psi_o$ will be:
\begin{equation}\label{singlmeas}
\psi_o(d\mu_1,d\mu_2)= (2\pi )^{-1} \delta (\mu_1- \mu_2)d\mu_1d\mu_2
\end{equation}
where $d\mu$ is the Lebesgue measure.

Finally the local single particle dynamics $P_{1,a}, P_{1,b}$, which
define the global dynamics through (\ref{locdyn}), will be (in Heisenberg
representation) of the form
\begin{equation}\label{p1p2ex}
P_{1,a}(f)(u):= T'_{1,a}(u) f(T_{1,a}u) \quad ,\quad
P_2(g)(v):= g(T_{2,b}v) T'_{2,b}(v)
\end{equation}
where $T_{1,a},T_{2,b}: S\to S$ are differentiable functions with
derivatives $T'_{1,a},T'_{2,b}$ which are strictly positive almost
everywhere and such that
$$\int^{2\pi}_0\int^{2\pi}_0\delta(\lambda_1-\lambda_2)T'_{1,a}
(\lambda_1)T'_{2,b}(\lambda_2){d\lambda_1d\lambda_2\over2\pi}=1$$
(cf. Section (4) below for an example of such functions). Notice that
the local dynamics (\ref{p1p2ex}) have the form (\ref{p1p2}).
In order to calculate the explicit form of the measure $\psi_o \circ P$,
corresponding to the choices (\ref{singlmeas}) of $\psi_o $
and (\ref{p1p2ex}) of $P$,  notice that, with these choices,
the correlation $\langle S^{(1)}_aS^{(2)}_b\rangle$
(cf. (\ref{eprcorsh})) becomes
\begin{equation}\label{eprcorhe2}
\int\int\tilde S^{(1)}_a(T_{1,a}\lambda_1)\tilde S^{(2)}_b(T_{2,b}\lambda_2)
\delta(\lambda_1-\lambda_2)
T'_{1,a}(\lambda_1)T'_{2,b}(\lambda_2){d\lambda_1d\lambda_2\over2\pi}
\end{equation}
and, with the notation
$$
\tilde S^{(j)}_x(T_{j,x}\lambda_j)=:S^{(j)}_x(\lambda_j)\ ;\quad
\quad j=1,2\ ;\quad x=a,b
$$
(notice that what we measure are the $\tilde S^{(j)}_x(T_{j,x}\lambda_j)$
and not the $\tilde S^{(j)}_x(\lambda_j)$) we finally obtain
\begin{equation}\label{finform}
\langle S^{(1)}_aS^{(2)}_b\rangle =(2\pi)^{-1}\int_0^{2\pi}
S^{(1)}_a(\lambda)S^{(2)}_b(\lambda)T'_{1,a}(\lambda)T'_{2,b}
(\lambda)d\lambda
\end{equation}

\medskip

\noindent{\bf Remark} Notice that, if $p_o(\lambda_1,\lambda_2)$ is any
probability density, then
$$\int\int\tilde S^{(1)}_a(T_{1,a}\lambda_1)\tilde S^{(2)}_b(T_{2,b}\lambda_2)
p_o(\lambda_1,\lambda_2)
T'_{1,a}(\lambda_1)T'_{2,b}(\lambda_2)d\lambda_1d\lambda_2=$$
$$=\int\!\!\int
\tilde S^{(1)}_a(T_{1,a}\lambda_1)\tilde S^{(2)}_b(T_{2,b}\lambda_2)
p_o(\lambda_1,\lambda_2)dT_{1,a}(\lambda_1)dT_{2,a}(\lambda_2)$$
Therefore, with the change of variables
$$
\mu_1=:T_{1,a}\lambda_1 \quad  ;\quad \mu_2=:T_{2,b}\lambda_2
$$
the correlations become:
$$\langle S^{(1)}_aS^{(2)}_b\rangle=
\int\int\tilde S^{(1)}_a(\mu_1)\tilde S^{(2)}_b(\mu_2)
p_o(T_{1,a}^{-1}\mu_1, T_{2,b}^{-1}\mu_2)d\mu_1d\mu_2$$
By taking limits the above identity continues to hold when $p_o$ is a
distribution. Therefore, if we choose
$ p_o(\mu_1, \mu_2)=(2\pi )^{-1}\delta(\mu_1 - \mu_2)$, then an equivalent
form for the correlations (\ref{eprcorhe2}) is
$$\langle S^{(1)}_aS^{(2)}_b\rangle=
\int\int\tilde S^{(1)}_a(\mu_1)\tilde S^{(2)}_b(\mu_2)
(2\pi )^{-1}\delta(T^{-1}_{1,a}\mu_1-T^{-1}_{2,b}\mu_2){d\mu_1d\mu_2
\over2\pi}$$
Which is the form originally used in \cite{[AcRe00b]}

\section{  Complete specification of the experiment }

Now let us make the following choices
for $T_{1,a}$, $T_{2,b}$, $\tilde S^{(1)}_a$, $\tilde S^{(2)}_b$:
\begin{equation}
T'_{2,b}(\lambda)=\sqrt{2\pi}\qquad (\hbox{constant})\label{dyn1}
\end{equation}
\begin{equation}
T'_{1,a}(\lambda)={\sqrt{2\pi}\over4}\,|\cos(\lambda-a)|\label{dyn2}
\end{equation}
$$S^{(1)}_x(\lambda)=\hbox{ sgn }(\cos(\lambda-x))$$
$$S^{(2)}_x=-S^{(1)}_x \qquad \hbox{the singlet condition}$$

With these choices, the correlations
$$\langle S^{(1)}_aS^{(2)}_b\rangle =
(2\pi)^{-1}\int S^{(1)}_a(\lambda)S^{(2)}_b(\lambda)T'_{1,a}(\lambda)T'_{2,b}
(\lambda)d\lambda$$
become
$$
=-\int^{2\pi}_0 
\hbox{sgn} (\cos(\lambda-a))\hbox{ sgn}(\cos(\lambda-b))
{1\over4}|\cos(\lambda-a)|d\lambda
$$
$$=\! - {1\over4}\int\hbox{sgn}(\cos(\lambda-b))\!\cos(\lambda-a)d\lambda=\!
- \cos(b-a)$$
which coincide with the singlet (EPR) correlations

\section{  Conclusions}

We have constructed:

(i)  A classical state space for a system composed of two particles
$(1,2)$  (the product of two copies of the interval $[0,2\pi]$)

(ii) a family  of local, classical, reduced Heisenberg dynamics, for the
system $(1,2)$, parametrized by the angles in the interval $[0,2\pi]$

(iii) for each of the two particles a family (also parametrized by the angles
in the interval $[0,2\pi]$) of classical binary observables (${\pm 1}$ valued
functions) denoted $\{\tilde S^{(1)}_a \}$ and $\{\tilde S^{(2)}_b\}$
respectively.

(iv) an initial distribution on the state space of the two particle system

\noindent with the following property: for any pair of angles $a,b$, in $[0,2\pi ]$,
the correlations of the two random variables $\tilde S^{(1)}_a,
\tilde S^{(2)}_b$, with respect to the measure $\psi_o$, evolved
with the local dynamics $P_{1,a}\otimes P_{2,b}$, are exactly the
singlet (EPR) correlations, i.e. $- \cos(b-a)$.

This means that we have built a local, realistic, hidden variable model
for the singlet (EPR) correlations.

This result reconciles the orthodox interpretation with physics, in the
sense that it is true that ``the act of measurement may determine the
value of some observable'', however this does not happen by virtue of
weird collapses, mysterious objectifications and bizarre nonlocalities,
but because of the quite natural and physically intuitive chameleon
effect. On the other hand it also reevaluates the hidden variable
theories by showing that they are not necessarily in contradiction with
locality and confirming the point of view, expressed in \cite{[Ac86]},
that the problem with hidden variable theories is not their existence
but their wild non uniqueness.

Moreover, since, because of locality, the values of all the observables of
particle $1$ can be calculated withouth knowing anything about particle $2$
and conversely, the above construction allows to realize the following
experiments.

Two, separate and non communicating, experimenters agree that one of
them will calculate only values of observables of particle $1$ and the
other only values of observables of particle $2$.

For each point $\lambda$ of the classical state space experimenter $1$ chooses,
arbitrarily and without communicating with the other experimenter,
an angle $a$ and computes the value of $\tilde S^{(1)}_a(\lambda)$.
The other experimenter does the same and computes $\tilde S^{(2)}_b(\lambda)$.

After a long (30 or 40 thousands) series of choices, the two
experimenters exchange their informations and compute the correlations
for all possible choices $(a,b)$: just as in the Eckert protocol for quantum
cryptography \cite{[Ek91]}.

Then they check a posteriori that, for appropriate choices of the angles
$(a,b)$, the Bell's inequalities are violated.

Another experiment, also realizable by the above construction, reproduces
the scheme adopted in the experiments to measure the violation of the Bell
inequality in quantum systems: two, separate and non communicating,
experimenters make the same agreement as above with the following variants:

-- they divide the input sequence of states $(\lambda_n)$ into four
subsequences (corresponding to four correlations), for example: the first
10.000 points, the second 10.000, and so on. This division is arbitrary,
but the four subsequences must be the same for the two experimenters.

-- each experimenter chooses, arbitrarily and without communicating with
the other one, two angles, let us say that $a,a'$ is chosen by $1$
and $b,b'$ by $2$

-- they agree a priori that $1$ will choose one of the two angles, say
$a$, in the $1$--st and $2$--d  subsequence of points and the other angle,
$a'$, in the $3$--d and $4$--th; while $2$ will choose one of the two angles,
say $b$, in the $1$--st and $3$--d  subsequence of points and the other angle,
$b'$, in the $2$--d and $4$--th.

After this, the two experimenters exchange their informations and
compute the correlations 
$$\langle a,b\rangle \quad ; \quad \langle a,b'\rangle \quad ; \quad
\langle a',b\rangle \quad ; \quad \langle a',b'\rangle $$
and they check that, for appropriate choices of the angles $(a,a',b,b')$,
the Clauser--Horn--Shimony--Holt form of the Bell's inequalities are violated.

\bigskip

\section{ Appendix: A dynamical theory of measurement }

In order to explain the emergence of local reduced dynamics, we summarize
here the dynamical theory of measurement proposed in (\cite{[Ac93]}).
This extends von Neumann's basic tenet that a good theory of measurement
should keep into account the joint system--apparatus evolution, by
introducing in it the notions of locality and causality.

All what said below, with appropriate interpretation of the symbols
involved, is equally valid both for classical and quantum systems.

\bigskip 
Let us introduce the following notations:

 $M$ measurement apparatus  

 $S$ system 

 $T^t_{(S,M)}$ joint interacting evolution 

 $T^t_S,T^t_M$ free evolutions  

 $(S_o,M_o)$ initial state

The basic postulate of the classical theory of measurement is that the 
properties of the system are not modified by the measurement process (at 
least before the measurement itself). This means that the 
perturbations due to the interaction with the measurement apparatus
are orders of magnitude less than the measured quantities. In formulae:
\begin{equation}
T^t_{S,M}(S_o,M_o) \ \hbox{ restricted to} \ S \ = \ T^t_SS_o
\label{C}\end{equation}
(for $t$ before the measurement). On the contrary the properties
of the measurement apparatus are modified due to the interaction with the 
system (otherwise there would be no measurement). In formulae:
\begin{equation}
T^t_{S,M}(S_o,M_o) \ \hbox{ restricted to} \ M \ \not= \ T^t_MM_o
\end{equation}
(for $t$ after the measurement). 

We speak of {\it chameleon effect } when condition (\ref{C}) is not justified.
Thinking of the interaction system--apparatus, one could say
that, in some sense, the surprising fact is that condition (\ref{C})  is a very
good approximation in many cases and not that it is violated in some 
other cases (e.g. quantum theory  or, among classical systems, chameleons).
\bigskip 
Consider now a system composed of two subsystems 
\begin{equation} S=(S_1,S_2)\ ;\quad M=(M_1,M_2)\end{equation}
Suppose that the two subsystems are spatially separated and subjected
each to a different measurement. Therefore their evolution shall depend 
on several interactions. Let us write:
\begin{equation}T^t_{(S,M)}=T^t_{(S_1,S_2,M_1,M_2)}\label{1}\end{equation}
We want to introduce the locality condition in the dynamical
evolution (\ref{1}). It is reasonable to expect that the most
general evolution (\ref{1}) is nonlocal
and that the class of local evolutions is very particular.
The solution of this problem, which shall be given in several steps, 
confirms this expectation.

-- i) It is convenient to use the Hamiltonian description of the dynamics:
\begin{equation}T^t_{(S,M)}=e^{-itH}\end{equation}
because in it the various contributions to the interaction are better 
separated. Let us write therefore
\begin{equation}H=H_S+H_M+H_I\end{equation}
separating the contribution of the system $H_S$, of the apparatus $H_M$, 
and of the interaction $H_I$.

-- ii) Since $S=(S_1,S_2)$ we shall have
\begin{equation}H_S=H_{S_1}+H_{S_2} +H_{S_1,S_2}\end{equation}
but the two systems are separated, therefore the {\bf first locality assumption}
is that their interaction is negligible, that is $H_{S_1,S_2}=0$ or, 
equivalently:
\begin{equation}H_S=H_{S_1}+H_{S_2}   \end{equation}

-- iii) Similarly $M=(M_1,M_2)$ and
\begin{equation}H_M=H_{M_1}+H_{M_2}+H_{M_1,M_2}\end{equation}
If, as in the EPR experiment, we suppose that the only constraint between the 
two measurements is their simultaneity, then we arrive to the {\bf second
locality assumption }, that is $H_{M_1,M_2}=0$ or, equivalently:
\begin{equation}H_M=H_{M_1}+H_{M_2}\end{equation}

-- iv) Eventually the interaction Hamiltonian shall have the form
\begin{equation}H_I=H_{S,M}=H_{S_1,S_2,M_1,M_2}=
H_{S_1,M_1} + H_{S_1,M_2} + H_{S_2,M_2} + H_{S_2,M_1}\end{equation}
and, again because of the spatial separation, it is natural to introduce the 
{\bf third locality assumption } in the form
\begin{equation}H_{S_1,M_2}=H_{S_2,M_1}=0\end{equation}
or, equivalently
\begin{equation}H_I=H_{S_1,M_1}+H_{S_2,M_2}\end{equation}
\bigskip

\noindent{\bf Definition}. A local dynamical law for the system
$(S_1,S_2,M_1,M_2)$, is given by an Hamiltonian of the form
\begin{equation}H=(H_{S_1}+H_{M_1}+H_{S_1,M_1})+(H_{S_2}+H_{M_2}+H_{S_2,M_2})
\label{2}\end{equation}
From this it follows that the locality condition (\ref{2})
is equivalent to the factorization condition
\begin{equation}T^t_{(S_1,S_2,M_1,M_2)}=T^t_{(S_1,M_1)}\otimes T^t_{(S_2,M_2)}
\label{3}\end{equation}
which corresponds to the {\bf factorization of the state space of the 
total system} given by
\begin{equation}{\cal H}_{S_1}\otimes{\cal H}_{S_2}\otimes{\cal H}_{M_1}\otimes{\cal
H}_{M_2}=\Bigl({\cal H}_{S_1}\otimes{\cal H}_{M_1}\Bigr)\otimes
\Bigl({\cal H}_{S_2}\otimes {\cal H}_{M_2}\Bigr)\label{4}\end{equation}
Let us now introduce the causality condition: at the initial instant, the
system cannot know which measurement shall be done on it. This means that the
initial preparation (state) of the system should be statistically 
independent on the state of the apparatus and mathematically it is expressed 
in the  form: 
\begin{equation}
\sigma_o=\hbox{ initial state of } \ (S_1,S_2,M_1,M_2)=
\sigma_{1,2} \otimes\sigma_{M_1,M_2}
\label{5}\end{equation}
where $\sigma_{S_1,S_2}$ is the initial state of $(S_1,S_2)$ and 
$\sigma_{M_1,M_2}$ the initial state of $(M_1,M_2)$. Notice that 
{\bf also causality is expressed by means of a factorization condition of
the state space of the total system}. However, being given by:
\begin{equation}{\cal H}_{S_1}\otimes{\cal H}_{S_2}\otimes{\cal H}_{M_1}\otimes
{\cal H}_{M_2}=\Bigl({\cal H}_{S_1}\otimes{\cal H}_{S_2}\Bigr)\otimes
\Bigl({\cal H}_{M_1}\otimes{\cal H}_{M_2}\Bigr)
\label{6}\end{equation}
this factorization is different from that of the dynamics.
Summing up: both locality and causality are expressed by a factorization 
property, however they correspond to different factorizazions of the 
state space.
\medskip
The chameleon effect is a corollary of this difference: the state 
at the instant of measurement does not factorize, that is in general for $t>0$,  
one cannot write it in the  form  
\begin{equation}\sigma(t)=T^t\sigma_o=\sigma_{a,b}(t)
\not=\sigma_{S_1,M_1}(t)\otimes\sigma_{S_2,M_2}(t)
\end{equation}
In particular, if the apparatus $M_1=M_a$ is predisposed for the measurement 
of $S^1_a$ and the apparatus $M_2=M_b$ for the measurement of $S^2_b$, then 
one shall have
\begin{equation}\label{st}
\sigma(t)=T^t_{(S,M)}\sigma_o=T^t\sigma_o=\sigma_{a,b}^{t}
\end{equation}
Notice that {\bf all what said up to now is equally valid both
for classical and for quantum sistems} (to obtain the latter ones it is 
sufficient to substitute the Liouvillian for the Hamiltonian and the 
Poisson brackets for the commutator). This shows that the chameleon effect 
is a general characteristic both of classical and quantum physics. 

Since the correlations between $S^1_a$ and $S^2_b$ are determined by the state
$\sigma_{a,b}^t$ through the formula
\begin{equation}\label{corrc}
E(S^1_aS^2_b)=\int S^1_a(x)S^2_b(x)\sigma_{a,b}^t(dx)=
\sigma_{a,b}^t(S^1_aS^2_b)
\end{equation}
in the classical theory, and through the formula 
\begin{equation}\label{corrq}
E(S^1_aS^2_b)=
\langle \sigma_{a,b}^t, S^1_a\otimes S^2_b\sigma_{a,b}^t\rangle=
\sigma_{a,b}^t(S^1_aS^2_b)
\end{equation}
in quantum theory, it follows that in both cases the expectation value, which 
determines the correlations, also depends on the pair $(a,b)$:
\begin{equation}E(S^1_aS^2_b)=E_{a,b}(S^1_aS^2_b)
\label{7}\end{equation}
and it is well known that, {\bf when the expectation value $E$ depends 
on $(a,b)$, it is impossibile to apply the Bell inequality}.

Finally let us remark that both (\ref{corrc}) and (\ref{corrq}) refer to
the joint (system, apparatus) system.
If we want that also the two apparata make local independent choices,
then there should not be correlations between their states. This means
that, in addition to the causality condition (\ref{5}), we must also
require the factorization
\begin{equation}\label{factapp}
\sigma_{M_a,M_b}= \sigma_{M_a}\otimes\sigma_{M_b}
\end{equation}
so that the initial state becomes:
\begin{equation}
\sigma_{a,b}
= \sigma_{1,2}\otimes\sigma_{M_a}\otimes\sigma_{M_b}
\end{equation}

Recalling now the factorization condition (\ref{3}) and the fact
that the symbol $S^1_aS^2_b$ is a short--hand notation for
\begin{equation}
S^1_a\otimes S^2_b\otimes 1_{M_a}\otimes 1_{M_b}
\end{equation}
and denoting $\overline \sigma_{M_a}$ (resp. $\overline \sigma_{M_b}$)
the partial expectation over the $M_a$--algebra (resp. $M_b$--algebra),
we see that both correlations (\ref{corrc}) and (\ref{corrq}) can be
rewritten in the form
\begin{equation}\label{reddyn}
\sigma_{a,b}^t(S^1_aS^2_b) =
(\sigma_{1,2}\otimes\sigma_{M_a}\otimes\sigma_{M_b}) (
T^t_{(1,2,M_a,M_b)}(S^1_aS^2_b) )=
\end{equation}
\begin{equation}
=\sigma_{1,2}  \left(
\overline \sigma_{M_a} [T^t_{(1,M_a)}(S^1_a\otimes 1_{M_a})]\otimes
\overline \sigma_{M_b} [T^t_{(2,M_b)}(S^2_b\otimes 1_{M_b})]
\right)
\end{equation}
Introducing the local reduced dynamics
\begin{equation}\label{locreddyn}
P^t_{1,a}(S^1_a):=\overline \sigma_{M_a} [T^t_{(1,M_a)}(S^1_a\otimes 1_{M_a})]
\quad ; \quad
P^t_{2,b}(S^2_b):=\overline \sigma_{M_b} [T^t_{(2,M_b)}(S^2_b\otimes 1_{M_b})]
\end{equation}
we finally obtain
\begin{equation}
\sigma_{a,b}^t(S^1_aS^2_b)
=\sigma_{1,2}  \left( P^t_{1,a}(S^1_a) \otimes P^t_{2,b}(S^2_b)
\right)
\end{equation}
Summing up we have proved that if: (i) the locality condition (\ref{3}),
(ii) the causality condition (\ref{5}), (iii) the local independence
condition (\ref{factapp}), are satisfied, then also the reduced dynamics
is local in the sense of (\ref{locdyn}).

Conversely, if we start from local reduced dynamics $P^t_{1,a},P^t_{2,b}$
then, by taking unitary dilations, we can always construct larger
(automorphic) dynamics $T^t_{(1,M_a)}$ and $T^t_{(2,M_b)}$ and therefore
a global (automorphic) dynamics which satisfies the locality condition
(\ref{3}). Therefore to start from reduced dynamics is not a
restriction.

Since, in the EPR type experiments, conditions (i), (ii), (iii) are
always satisfied, the above arguments justify our use of local reduced
dynamics. ÿ

In conclusion: we speak of ``chameleon effect'' when:

-- i) the locality condition is satisfied

-- ii) the causality condition is satisfied

-- iii) the state of the total system can depend on the joint measurement
$(a,b)$ ($T^t_{a,b}\sigma_o=\sigma_{a,b}^t$)

in these cases, since the state depends on the joint measurement, 
the Bell inequality is violated. Therefore the Bell inequality can be 
violated in full respect of locality and causality.

For those systems for which the application of the postulate (\ref{C}) of
classical measurement is justified one can apply the usual statistics.

Chameleons provide a simple example of classical system in which the above
postulate is not  justified. Therefore it is natural to expect that the 
statistics of the color of a set of pairs of chameleons will be
different from that of a set of pairs of balls. The latter obey 
classical statistics (and the Bell inequality is satisfied), while
our experiment proves that a set of pairs of ``chameleon--like
entities''
can find an agreement on how to distribute their answers to binary questions
(corresponding to $S^1_a, S^1_b,S^1_c, \dots$) so to reproduce the EPR
correlations hence violate Bell's inequality. It is conjectured that,
by the above described chameleon effect, an arbitrary set of correlations
for random variables with values in the interval $[-1,1]$ can be obtained.
However, at the moment, this is an open problem.

For these reasons, just as balls and dice are natural symbols for classical
statistics, chameleons could be natural candidates to become the symbol of 
quantum statistics.

\end{document}